\documentclass[12pt,preprint]{aastex}
\usepackage{natbib}
\usepackage{graphicx}
\shorttitle{Constraining annihilating dark matter}
\begin{document}
\title{Constraining annihilating dark matter by radio continuum spectrum of the Large Magellanic Cloud}
\author{Man Ho Chan \& Chak Man Lee}
\affil{Department of Science and Environmental Studies, The Education University of Hong Kong, Hong Kong, China}
\email{chanmh@eduhk.hk}

\begin{abstract}
Recent radio observations have obtained stringent constraints for annihilating dark matter. In this article, we use the radio continuum spectral data of the Large Magellanic Cloud (LMC) to analyze the dark matter annihilation signals. We have discovered a slightly positive signal of dark matter annihilation with a $1.5\sigma$ statistical significance. The overall best-fit dark matter mass is $m_{\rm DM} \approx 90$ GeV, annihilating via $b\bar{b}$ channel. We have also constrained the $3\sigma$ lower limits of dark matter mass with the standard thermal dark matter annihilation cross section for the $e^+e^-$, $\mu^+\mu^-$, $\tau^+\tau^-$ and $b\bar{b}$ channels. 
\end{abstract}


\section{Introduction}
Dark matter is one of the most mysterious components in our universe. Although observations of galaxies, galaxy clusters and the Cosmic Microwave Background reveal the existence of dark matter, we don't know very much about the nature of dark matter. Some benchmark models suggest that dark matter can self-annihilate to give high-energy particles, such as electrons, positrons, neutrinos and photons \citep{Roszkowski}. In view of this property, many telescopes are trying to detect dark matter annihilation signals directly, such as the Alpha Magnetic Spectrometer (AMS) \citep{Aguilar}, DAMPE \citep{Ambrosi} and Fermi-LAT \citep{Ackermann,Albert}. However, no promising signal has been detected.

On the other hand, radio observations can also contribute to detect dark matter annihilation. The high-energy electrons and positrons produced in dark matter annihilation can emit synchrotron radiation in radio bands when there is a strong magnetic field. Many previous studies have attempted to detect or constrain the signal of dark matter annihilation through radio analyses. In particular, most of the studies have focused on large galaxies \citep{Egorov,Chan2} or galaxy clusters \citep{Colafrancesco,Storm,Colafrancesco2,Chan3,Chan4}. Only a few studies have focused on dwarf galaxies \citep{Chan5,Kar,Chan6} because the magnetic field strength in dwarf galaxies is relatively small so that the synchrotron signals are weaker. Also, the diffusion of the high-energy electrons and positrons is more efficient, which would suppress the signals of dark matter annihilation. 

However, using nearby dwarf galaxies to analyze dark matter annihilation signals has some advantages. First of all, some dwarf galaxies are dark matter-dominated so that relatively larger signals of dark matter annihilation would be potentially detected. Moreover, the baryonic content in dwarf galaxies is usually smaller, which might give a less significant baryonic radio contribution. Recent studies using the radio continuum spectral analysis can effectively differentiate the radio contribution of dark matter from the total radio spectrum, which can give tighter constraints for dark matter \citep{Chan3,Chan6,Chan4}. If we can find some nearby dwarf galaxies which have relatively large magnetic field strength, some better constraints could be obtained by following the spectral analysis.

In this article, we show that the radio continuum spectrum of a nearby dwarf galaxy, the Large Magellanic Cloud (LMC), might be able to give some hints of dark matter signals. Although the statistical significance of the dark matter annihilation signal is not very large, we can get some stringent lower bounds of dark matter mass based on the analysis. 

\section{The dark matter annihilation model}
Through dark matter annihilation, high-energy electrons and positrons would be produced and then cooled down gradually during diffusion. The diffusion and cooling of electrons and positrons can be overall governed by the following diffusion-cooling equation \citep{Ginzburg,Atoyan}
\begin{eqnarray}
\frac{\partial}{\partial t}\frac{dn_{\rm e}}{dE}&=&\frac{D(E)}{r^2}\frac{\partial}{\partial r}\left(r^2\frac{\partial }{\partial r}\frac{dn_{\rm e}}{dE}\right)+\frac{\partial}{\partial E}\left[b_{\rm T}(E,r)\frac{dn_{\rm e}}{dE}\right]\nonumber\\
&&+Q(E,r),
\label{diffusion}
\end{eqnarray}
where $dn_e/dE$ is the electron/positron density spectrum, $D(E)$ is the diffusion function, $b_{\rm T}(E,r)$ is the cooling rate function, and $Q(E,r)$ is the source density spectrum from dark matter annihilation. The diffusion function is usually written in terms of an energy-dependent function $D(E)=D_0(E/1\;{\rm GeV})^{\delta}$, where $D_0$ is the diffusion coefficient and $\delta$ is the diffusion index. In the followings, we take a benchmark value of $\delta=1/3$ \citep{Kolmogorov}, which has been assumed in many related studies \citep{Ackermann2}.

The cooling rate of electrons and positrons is dominated by the synchrotron emission and the inverse Compton scattering of the background photon field. The other cooling processes such as Bremsstrahlung loss are nearly negligible in dwarf galaxies. The cooling rate function (in the unit of $10^{-16}$ GeV s$^{-1}$) is expressed by \citep{Siffert}
\begin{equation}
b_{\rm T}(E,r)=0.0254E^2B^2 + 1.02U_{\rm rad}(r)E^2,
\label{cooling}
\end{equation}
where $B$ is the magnetic field strength (in $\mu$G) of the galaxy, $E$ is the energy (in GeV) of an electron or positron and $U_{\rm rad}$ is the energy density of the Interstellar Radiation Field (ISRF) in the unit of ${\rm eV~cm}^{-3}$.

In Eq. (\ref{diffusion}), the particle-injection source term is given by  \citep{Vollmann}
\begin{equation}
Q(E,r)=\frac{\langle \sigma v \rangle [\rho_{\rm DM}(r)]^2}{2m_{\rm DM}^2}\frac{dN_{\rm e,inj}}{dE},
\end{equation}
where $\rho_{\rm DM}(r)$ is the dark matter density profile and $dN_{\rm e,inj}/dE$ is the injected energy spectrum of dark matter annihilation. The injected energy spectrum depends on the annihilation channels, which can be obtained in \citet{Cirelli}. Here, we take the thermal relic dark matter annihilation cross section predicted by standard cosmology: $\langle \sigma v \rangle= 2.2\times 10^{-26} {\rm cm}^3~{\rm s}^{-1}$ \citep{Steigman}. Assuming in equilibrium condition, we have $\frac{\partial}{\partial t}(\frac{dn_{\rm e}}{dE})=0$ in Eq. (\ref{diffusion}). Putting the boundary condition $\frac{dn_{\rm e}(r_{\rm h},E)}{dE}=0$ with diffusion halo radius $r_{\rm h}$, the general solution of the equilibrium electron density spectrum can be obtained in terms of the Fourier-series representation of the Green's function \citep{Vollmann}:
\begin{eqnarray}
\frac{dn_{\rm e}}{dE}(E,r)&=&\sum_{n=1}^{\infty}\frac{2}{b_{\rm T}(E,r)r_{\rm h}}\frac{\sin\left(\frac{n\pi r}{r_{\rm h}}\right)}{r}
\nonumber\\
&&\times\int_E^{^{m_{\rm DM}}}dE'e^{-n^2[\eta(E)-\eta(E')]}\nonumber\\
&&\times\int_0^{r_{\rm h}}dr'r'\sin\left(\frac{n\pi r'}{r_{\rm h}}\right)Q(E',r'),
\label{dndE}
\end{eqnarray}
where the dimensionless variable $\eta(E)$ is given by
\begin{eqnarray}
\eta(E)&=&\frac{1}{1-\delta}\left(\frac{6.42\pi\;{\rm kpc}}{r_{\rm h}}\right)^2\left(\frac{D_0}{10^{28}{\rm cm^2/s}}\right)
\nonumber\\
&&\times\left(\frac{1}{1+\left({B}/{3.135 \;\mu{\rm G}}\right)^2}\right)\left(\frac{1\;{\rm GeV}}{E}\right)^{1-\delta}.
\end{eqnarray}
Owing to the presence of the exponential factor in Eq. (\ref{dndE}), for $\delta<1$, the convergence of the series will be effectively achieved by only taking the first certain number of summation terms ($n=40$ for the present study).

The average power at frequency $\nu$ under magnetic field $B$ for synchrotron emission induced by the dark matter annihilation is given by \citep{Longair,Storm}
\begin{equation}
P_{\rm syn}(\nu)=\int_0^{\pi}d\theta\frac{(\sin\theta)^2}{2}2\pi\sqrt{3}r_{\rm e}m_{\rm e}c\nu_{\rm g}F_{\rm syn}\left(\frac{x}{\sin\theta}\right),
\end{equation}
where $\nu_{\rm g}=eB/(2\pi m_{\rm e}c)$, $r_{\rm e}$ is the classical electron radius, and $F_{\rm syn}(x/\sin\theta)=(x/\sin\theta) \int_{x/\sin\theta}^{\infty}K_{5/3}(u)du$. The quantity $x$ is defined as
\begin{equation}
x=\frac{2\nu}{3\nu_{\rm g}\gamma^2}\left[1+\left(\frac{\gamma\nu_{\rm p}}{\nu}\right)^2\right]^{3/2},
\end{equation}
where $\gamma$ is the Lorentz factor of the electrons/positrons, and $\nu_{\rm p}=8890[n(r)/1\;{\rm cm}^{-3}]^{1/2}$ Hz is the plasma frequency with the number density of the thermal electrons $n(r)\sim 1\;{\rm cm^{-3}}$. For dwarf galaxies, the profiles for both magnetic field and radiation density are in disc shape with thickness $\bar{z}_0$. The radio flux emitted within a solid angle $\Delta \Omega$ by a dwarf galaxy due to dark matter annihilation, with a distance $D_{\rm L}$ as observed from Earth, is finally expressed in cylindrical coordinates $(R,z)$ with azimuthal symmetry as
\begin{equation}
S_{\rm DM}(\nu)=
2\frac{\Delta\Omega}{4\pi}\int_{s_{\rm min}}^{s_{\rm max}} ds\int_{m_{\rm e}}^{m_{\rm DM}}P_{\rm syn}(\nu)\frac{dn_{\rm e}}{dE}dE,\;
\label{Sradio}
\end{equation}
where
\begin{equation}
\Delta\Omega=2\pi\int_0^{\bar{\theta}_{\rm max}}\sin\bar{\theta} (2d\bar{\theta}),
\end{equation}
\begin{equation}
\bar{\theta}_{\rm max}\approx \arctan\left(\frac{r_{\rm h}}{D_{\rm L}}\right),
\end{equation}
\begin{equation}
s_{\rm min}=\frac{ D_{\rm L} -\frac{\bar{z}_0}{2}  }{\cos\bar{\theta}},\,\,\,s_{\rm max}=\frac{ D_{\rm L}+ \frac{\bar{z}_0}{2}  }{\cos\bar{\theta}},
\end{equation}
\begin{equation}
R=D_{\rm L}\tan\bar{\theta},\,\,\,z=s\cos\bar{\theta}-D_{\rm L}.
\end{equation}
In Eq. (\ref{Sradio}), the factor 2 indicates the contributions of both high-energy electrons and positrons, and $s$ is the line-of-sight distance which is defined as $r=\sqrt{D_{\rm L}^2+s^2-2D_{\rm L}s\cos\bar{\theta}} $ (see Fig.~1 for the schematic diagram of the geometry considered). 

\section{Results}
We use the LMC as our target to constrain dark matter. Assume that the dark matter density profile of the LMC is spherically symmetric. It can be best-described by a modified Navarro-Frenk-White (NFW) profile (included the effect of tidal stripping), which can be expressed as \citep{Hayashi}
\begin{equation}
\rho_{DM}(r)=\frac{\rho_0}{1+\left(\frac{r}{r_0}\right)^3}\left[\frac{1}{\frac{r}{r_{0,{\rm NFW}}}\left(1+\frac{r}{r_{0,{\rm NFW}}}\right)^2}\right]
\end{equation}
where the parameters $\rho_0=(8.16\pm0.30)\times10^{6}\; {\rm M_{\odot}}\;{\rm kpc}^{-3}$, $r_0=6.36\pm2.08\; {\rm kpc}$, and $r_{0,{\rm NFW}} = 9.04\pm2.43\; {\rm kpc}$ are assumed for the LMC \citep{Siffert}. The modified NFW profile can give an excellent fit to the observed rotation curve of the LMC \citep{Siffert}. Therefore, the systematic uncertainty using this dark matter density profile is relatively small. The overall effect of this uncertainty will be discussed below. The distance to the galaxy $D_{\rm L}$ is 50 kpc \citep{Alves} and we take the value of $r_h \approx 8$ kpc equal to the size of the LMC galaxy.

For the radiation density of the LMC, we can write as \citep{Siffert}
\begin{equation}
U_{\rm rad}(R,z)=U_{\rm rad}^{\rm disc}(R)\exp(-|z|/h_0)
\end{equation}
with
\begin{equation}
h_0(R)=h_0(0)\exp(R/\xi),
\end{equation}
where $h_0(0)=0.14\;{\rm kpc}$ and $\xi=2.24\;{\rm kpc}$. Here, the disc thickness of the LMC is not a constant. Following the kinematics of carbon stars, the thickness of the LMC increases radially from its center \citep{Alves1, Alves}, and can be parameterized linearly in the unit of kpc by \citep{Van der Marel}
\begin{equation}
\bar{z}_0=0.27 + 1.23 \left(\frac{D_{\rm L}\tan\bar{\theta}}{5.5\, {\rm kpc}}\right).
\end{equation}
The radial dependence of $U_{\rm rad}^{\rm disc}(R)$ is shown in Table 1 \citep{Siffert}. For the magnetic field strength, it varies with different regions inside the LMC \citep{Siffert}. It does not have a clear radial dependence and we have taken a more conservative average value $B=4.3$ $\mu$G derived from rotation measures to model the magnetic field strength \citep{Gaensler}. This value is within the reasonable range of the magnetic field strength in LMC \citep{Regis}.

As shown in previous studies, better constraints could be obtained if we can differentiate the dark matter contribution from the total radio emission \citep{Chan3,Chan6,Chan4}. First of all, the total radio flux density $S(\nu)$ can be written in the sum of two parts: thermal part $S_{\rm th}(\nu)$ and non-thermal part $S_{\rm nth}(\nu)$. The thermal part can be described by the following form \citep{Tasitsiomi}
\begin{equation}
S_{\rm th}(\nu)=S_{\rm th,0}(\nu/{\rm Hz})^{-0.1},
\end{equation}
with $S_{\rm th,0}=2.2\times10^3$ Jy.
For the non-thermal part, it consists of two components: dark matter annihilation $S_{\rm DM}$ and the background cosmic-ray contribution $S_{\rm CR}$. The former one $S_{\rm DM}$ is calculated by Eq. (\ref{Sradio}) while the latter one can be described by a simple power-law form $S_{\rm CR}\propto\nu^{-\alpha_{\rm CR}}$. The non-thermal radio flux density emitted by a galaxy can be written explicitly as 
\begin{equation}
S_{\rm nth}(\nu)=S_{\rm DM}(\nu)+S_{\rm CR,0}\left(\frac{\nu}{\rm GHz}\right)^{-\alpha_{\rm CR}}.
\end{equation}
In fact, the power-law form is the simplest model to describe the cosmic-ray contribution. This is predicted by some theoretical models (e.g. secondary emission model) \citep{Dennison} and simulations \citep{Nava}. The power-law model is also consistent with the observational data of many dwarf galaxies \citep{Srivastava}. There are some possible models (e.g. insitu model) which predict other spectral shapes, especially for describing the cosmic-ray spectra in galaxy clusters \citep{Schlickeiser,Thierbach}. We will also test these spectral functions in our analysis. 

Using the radio continuum spectrum obtained in \citet{Haynes} (see Table 2), we can determine the best-fit scenarios. The goodness of fits can be determined by the $\chi^2$ value, which is defined as
\begin{equation}
\chi^2=\sum_i \frac{[S(\nu)-S_i]^2}{\sigma_i^2},
\end{equation}
where $S_i$ is the observed radio flux and $\sigma_i$ is the $1\sigma$ uncertainty of the data. To fit the observed radio spectrum, we have four free parameters \{ $D_0$, $m_{\rm DM}$, $S_{\rm CR,0}$, $\alpha_{\rm CR}$ \} for each annihilation channel. For each annihilation channel with fixed $m_{\rm DM}$ and $D_0$, we can get the best-fit $S_{\rm CR,0}$ and $\alpha_{\rm CR}$ by minimizing the $\chi^2$ values. Here, we only consider a physical range of $D_0=10^{26}-10^{30}$ cm$^2$ s$^{-1}$ for dwarf galaxies. 

In Fig.~2, we plot the $\chi^2$ as a function of $m_{\rm DM}$ for 3 different values of $D_0$. When $m_{\rm DM}$ is very large, $S_{\rm DM}$ would tend to zero, which defines our null hypothesis (i.e. no dark matter annihilation). In this case, the entire radio spectrum is represented by the background cosmic-ray contribution $S_{\rm CR}$ only with $\chi_{\rm null}^2=19.1$. In Fig.~2, we can see that there exist some values of $m_{\rm DM}$ in which the $\chi^2$ values are smaller than $\chi_{\rm null}^2$. These represent positive signals of dark matter annihilation. However, only $b\bar{b}$ channel and $W^+W^-$ channel can give positive signals with statistical significance larger than $1\sigma$ ($\chi_{\rm null}^2-\chi^2>2.3$ for 2 extra degrees of freedom). The largest statistical significance of the signals corresponds to $m_{\rm DM}=90$ GeV via the $b\bar{b}$ channel with $D_0 \sim 10^{28}$ cm$^2$ s$^{-1}$ (with $1.51\sigma$ statistical significance). Therefore, only a weak signal of dark matter annihilation could be identified in the analysis. We show some of the best-fit parameters in Table 3 and plot the overall best-fit radio spectrum in Fig.~3. 

Moreover, we have examined the effects of two major uncertainties involved in the study: 1. the uncertainties of the dark matter density parameters $\rho_0$, $r_0$ and $r_{0,\rm NFW}$, and 2. the uncertainties of the magnetic field strength $B$. We consider the $1\sigma$ uncertainties of the density parameters and a wider range of magnetic field strength $B=2-8$ $\mu$G to see how they affect our results (taking $D_0=10^{28}$ cm$^2$ s$^{-1}$). In Figs.~4 and 5, we can see that the $b\bar{b}$ and $W^+W^-$ channels also give the best-fit $m_{\rm DM}$ with $\chi^2 \le 16.8$ ($\ge 1\sigma$ statistical significance), although the best-fit values of $m_{\rm DM}$ would change (see Table 4 for the best-fit ranges). Therefore, positive signals still remain even we have included the possible uncertainties. We have also tested some other spectral functions \citep{Thierbach} to model the cosmic-ray contributions. Nevertheless, these models give slightly smaller likelihoods compared with the power-law form. Generally speaking, the best-fit $m_{\rm DM}$ is consistent with the $m_{\rm DM}$ range from recent anti-proton analysis ($m_{\rm DM} \approx 64-88$ GeV) \citep{Cholis}. 

Apart from the best-fit scenarios, we can also obtain the minimum $m_{\rm DM}$ allowed in our analysis. If the $\chi^2$ value is too large for a certain value of $m_{\rm DM}$ such that it deviates from $\chi_{\rm null}^2$ by a $3\sigma$ range (99.73\% C.L.), that $m_{\rm DM}$ would be the $3\sigma$ minimum allowed dark matter mass $m_{\rm DM,min}$ for the annihilation dark matter model. For $D_0=10^{26}-10^{30}$ cm$^2$ s$^{-1}$, we can see that a larger $D_0$ can give a smaller value of $m_{\rm DM,min}$. When a conservative value of $D_0=10^{30}$ cm$^2$ s$^{-1}$ is taken, the smallest values of $m_{\rm DM,min}$ are approximately 25 GeV, 50 GeV, 40 GeV and 60 GeV respectively for the $e^+e^-$, $\mu^+\mu^-$, $\tau^+\tau^-$ and $b\bar{b}$ channels. There is no minimum allowed dark matter mass for the $W^+W^-$ channel based on this analysis apart from the threshold mass \footnote{Based on the Standard Model, the threshold mass of dark matter annihilating via the $W^+W^-$ channel is 80.4 GeV.}. Generally speaking, these are the conservative lower limits of the annihilating dark matter mass. In fact, the actual diffusion coefficient for dwarf galaxies like the LMC should be close to $D_0 \sim 10^{27}-10^{28}$ cm$^2$ s$^{-1}$ as theoretical models predict $D_0 \sim LV$ \citep{Jeltema}, where $L$ and $V$ are the injection scale and the turbulent velocity respectively. For a typical dwarf galaxy, we should have $L \sim 100$ pc and $V \sim 50$ km/s, which give $D_0 \sim 10^{27}-10^{28}$ cm$^2$ s$^{-1}$. Therefore, the lower limits of $m_{\rm DM}$ reported here are underestimated.

\section{Discussion}
In this article, we use the radio continuum spectrum of the LMC dwarf galaxy to determine the best-fit annihilating dark matter mass and constrain the minimum allowed annihilating dark matter mass. However, only a slightly positive signal of dark matter annihilation ($\approx 1.5 \sigma$) can be identified. The best-fit mass $m_{\rm DM}=90$ GeV is consistent with the recent claim based on the anti-proton analysis in \citet{Cholis}. Furthermore, we can obtain some stringent lower limits of dark matter mass for different popular annihilation channels. Although we have considered a very conservative diffusion coefficient $D_0=10^{30}$ cm$^2$ s$^{-1}$, our $3\sigma$ limits are still more stringent than the robust conservative limits obtained in \citet{Egorov} using the radio data of the M31 galaxy. Therefore, using radio data of dwarf galaxies might also be good for constraining annihilating dark matter. Note that we did not consider any boost factor in our analysis. Many stringent radio constraints obtained previously have considered the boost factor \citep{Chan5,Chan6} so that larger values of $m_{\rm DM,min}$ would be resulted. In order to minimize the systematic uncertainty and get more conservative constraints, we neglect the consideration of the boost factor. 

In our analysis, one important uncertainty is the diffusion coefficient $D_0$. Here, we set it to be a free parameter for fittings as it is very difficult for us to determine it independently from other observations. Nevertheless, if one can get a certain constraint on $D_0$ based on cosmic-ray analysis, some better constraints of $m_{\rm DM}$ could be obtained. Beside the diffusion coefficient, the magnetic field strength is another parameter which may have some uncertainty. Here, we have adopted a constant average magnetic field $B=4.3$ $\mu$G which is obtained in \citet{Gaensler}. Generally speaking, the morphology of magnetic field in a dwarf galaxy is usually very complicated and asymmetrical. It is quite difficult for us to express the magnetic field strength in terms of elementary functions with a few parameters. Nevertheless, the fluctuation of the magnetic field strength in the LMC is not very large \citep{Gaensler}. Therefore, using an average magnetic field strength in our analysis would be good enough to achieve our objectives. By considering a conservative range of $B=2-8$ $\mu$G, as seen in Fig.~5, we can obtain a wider possible range of the best-fit $m_{\rm DM}$. Furthermore, we have also tried some other functional forms of the cosmic-ray contribution such as the insitu model \citep{Schlickeiser,Thierbach} and the Rephaeli model \citep{Rephaeli}. However, no better fits could be obtained compared with the simple power-law form used.

Generally speaking, nearby dwarf galaxies are good targets for constraining dark matter because many of them are dark matter-dominated. Using the data of dwarf galaxies, previous gamma-ray studies have obtained stringent constraints for annihilating dark matter \citep{Ackermann,Albert}. In view of this, some previous studies have collected low-frequency ($<1$ GHz) radio data of nearby dwarf spheroidal galaxies to constrain dark matter \citep{Kar}. The constraints obtained are quite stringent for some of the annihilation channels. In particular, one recent study observing the LMC using Australian Square Kilometre Array Pathfinder (ASKAP) with a single frequency at 888 MHz has obtained some stringent lower limits of annihilating dark matter mass for the thermal relic annihilation cross section \citep{Regis}. We anticipate that more radio observations of nearby dwarf galaxies using state-of-the-art radio telescopes with different frequencies can better detect or constrain annihilating dark matter. 

\begin{table}
\caption{The radial energy density profile of the Interstellar Radiation Field (ISRF) around the disc of the LMC \citep{Siffert}. Note that we have taken the average value for the data at $R$ = 1.71 kpc shown in \citet{Siffert}.}
\begin{tabular}{ |c|c|}
\hline
$R$ (kpc)     & $U_{\rm rad}^{\rm disc}(R)$(eV cm$^{-3})$\\
\hline
0.56               &1.508              \\
1.21               &1.346              \\
1.33               &1.098              \\
1.36               &1.030              \\
1.51               &1.207              \\
1.71               &0.787              \\
1.72               &0.857              \\
2.17               &0.921              \\
2.21               &0.997              \\
2.73               &1.266              \\
2.78               &1.233              \\
3.67               &0.637              \\
3.95               &0.569              \\
4.19               &0.643              \\
4.32               &1.106              \\
4.79               &0.985              \\
5.77               &0.601              \\
5.85               &0.694              \\
6.07               &0.900              \\
6.56               &0.878              \\
7.64               &0.494              \\
7.92               &0.461              \\
\hline
\end{tabular}
\end{table}

\begin{table}
\caption{Integrated radio flux densities of the LMC \citep{Haynes}.}
\begin{tabular}{ |c|c|c|c|}
 \hline
$\nu ({\rm GHz})$  &$S(\nu)({\rm Jy})$ & $S_{\rm nth}(\nu)({\rm Jy})$   & Uncertainties(Jy)   \\
\hline
0.0197       & 5270           & 4860        & 1054                \\
0.045        & 2997           & 2619        & 450                 \\
0.0855       & 3689           & 3335        & 400                 \\
0.0968       & 2839           & 2489        & 600                 \\
0.158        & 1736           & 1403        & 490                 \\
0.408        & 925            & 622         & 30                  \\
1.4          & 529            & 261         & 30                  \\
2.3          & 412            & 157         & 50                  \\
2.45         & 390            & 137         & 20                  \\
4.75         & 363            & 126         & 30                  \\
8.55         & 270            & 47          & 35                  \\
\hline
\end{tabular}
\end{table}

\begin{table}
\caption{Best-fit parameters for the null hypothesis and the dark matter (DM) hypothesis for each annihilation channel.}

\begin{tabular}{ |l|c|c|c|c|c|}
 \hline
   &  $m_{\rm DM}$ & $S_0$ & $\alpha_{\rm CR}$ & $D_0$ & $\chi^2$ \\
   &  (GeV) & (Jy) & & (cm$^2$ s$^{-1}$) & \\
  \hline
  Null &   & 329 & 0.74 & & 19.13 \\
  \hline
  DM ($\tau^+\tau^-$) & 50 & 153 & 0.89 & $10^{30}$ & 18.78 \\
  DM ($b\bar{b}$) & 90 & 65 & 0.06 & $10^{28}$ & 14.97 \\
  DM ($W^+W^-$) & 120 & 60 & 0.30 & $10^{27}$ & 15.07 \\
 \hline
\end{tabular}
\end{table}

\begin{table}
\caption{Best-fit ranges of $m_{\rm DM}$ after considering the uncertainties of the dark matter density parameters and the magnetic field strength. Assume $D_0=10^{28}$ cm$^2$ s$^{-1}$.}

\begin{tabular}{ |l|c|c|}
 \hline
  Uncertainties &  Channel & best-fit $m_{\rm DM}$ (GeV) \\
  \hline
  Dark matter density & $b\bar{b}$ & 60-280 \\
                      & $W^+W^-$ & 90-110 \\
  \hline
  Magnetic field & $b\bar{b}$ & 50-220 \\
                 & $W^+W^-$ & 90-110 \\
 \hline
\end{tabular}
\end{table}

\begin{figure}
\begin{center}
\includegraphics[width=140mm]{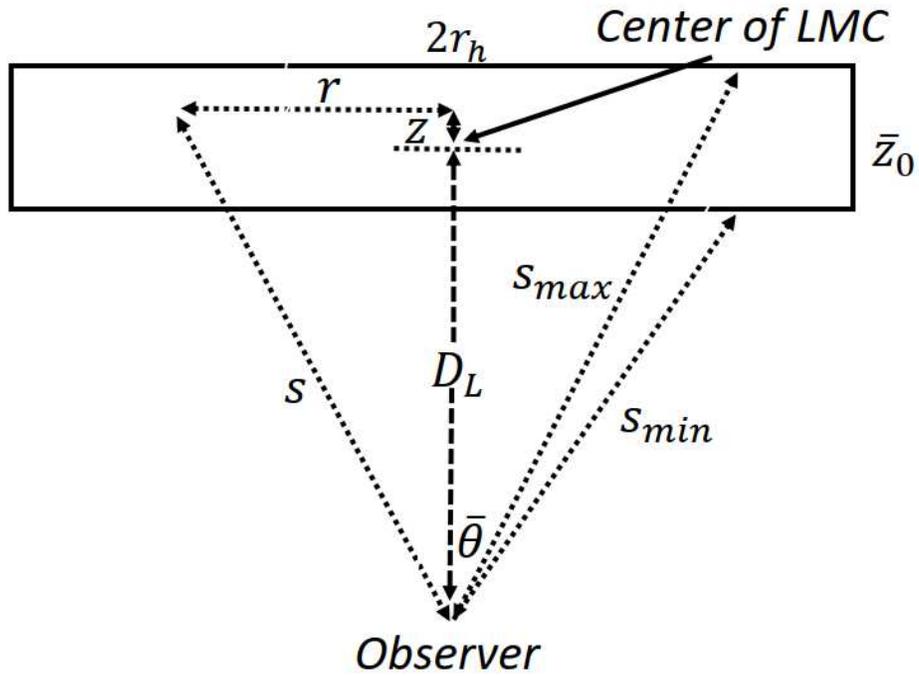}
\caption{A schematic diagram of the geometry of the LMC (not in scale). Here, the rectangle is the side-view cross section of the galactic disc of the LMC (the circular disc plane is facing the observer). } 
\label{Fig1}
\end{center}
\end{figure}

\begin{figure}
\begin{center}
\includegraphics[width=140mm]{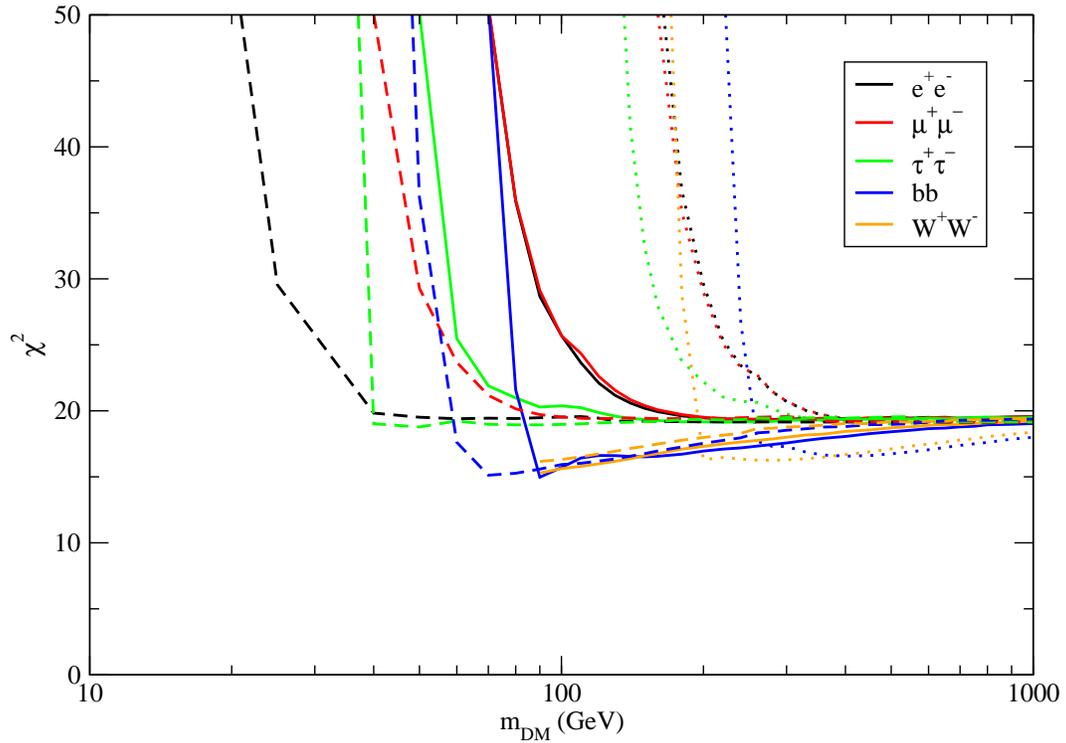}
\caption{The values of $\chi^2$ against $m_{DM}$ for different annihilation channels for 3 different values of the diffusion coefficient (dotted lines: $D_0=10^{26}$ cm$^2$ s$^{-1}$; solid lines: $D_0=10^{28}$ cm$^2$ s$^{-1}$; dashed lines: $D_0=10^{30}$ cm$^2$ s$^{-1}$).}
\label{Fig2}
\end{center}
\end{figure}

\begin{figure}
\begin{center}
\includegraphics[width=140mm]{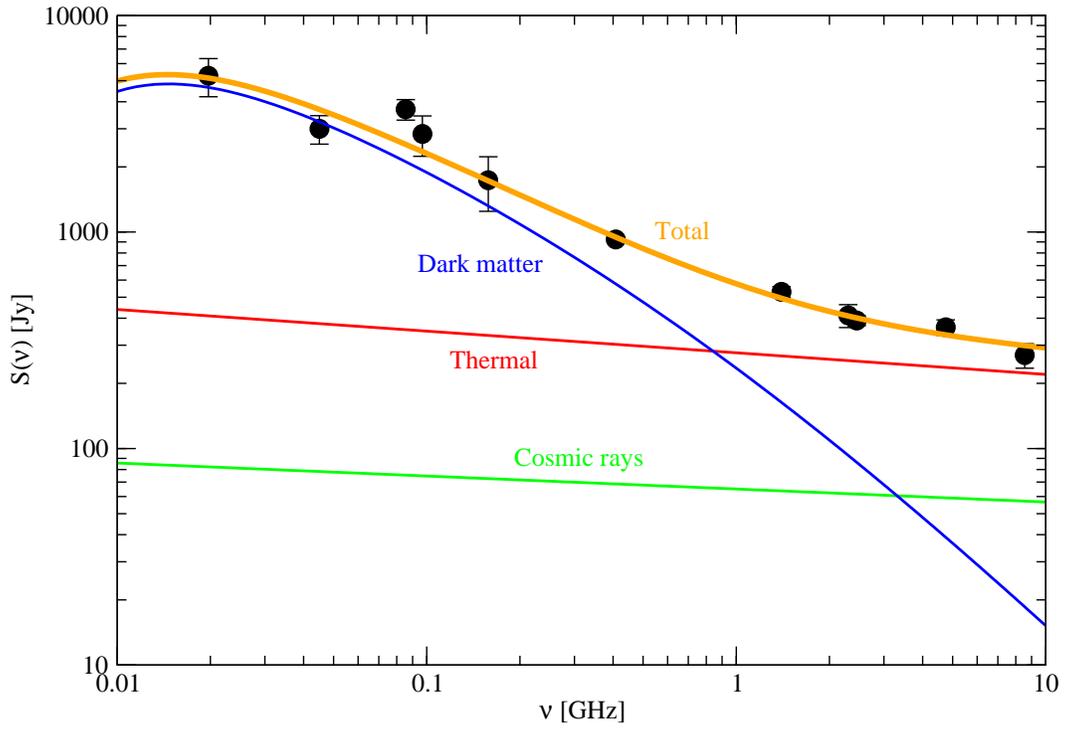}
\caption{The orange solid line indicates the overall best-fit radio continuum spectrum ($b\bar{b}$ channel with $m_{DM}=90$ GeV and $D_0=10^{28}$ cm$^2$ s$^{-1}$). The red line, green line and blue line represent the contributions of the thermal, best-fit cosmic-ray and best-fit dark matter components respectively. The data with error bars are extracted from \citet{Haynes}.}
\label{Fig3}
\end{center}
\end{figure}

\begin{figure}
\begin{center}
\includegraphics[width=140mm]{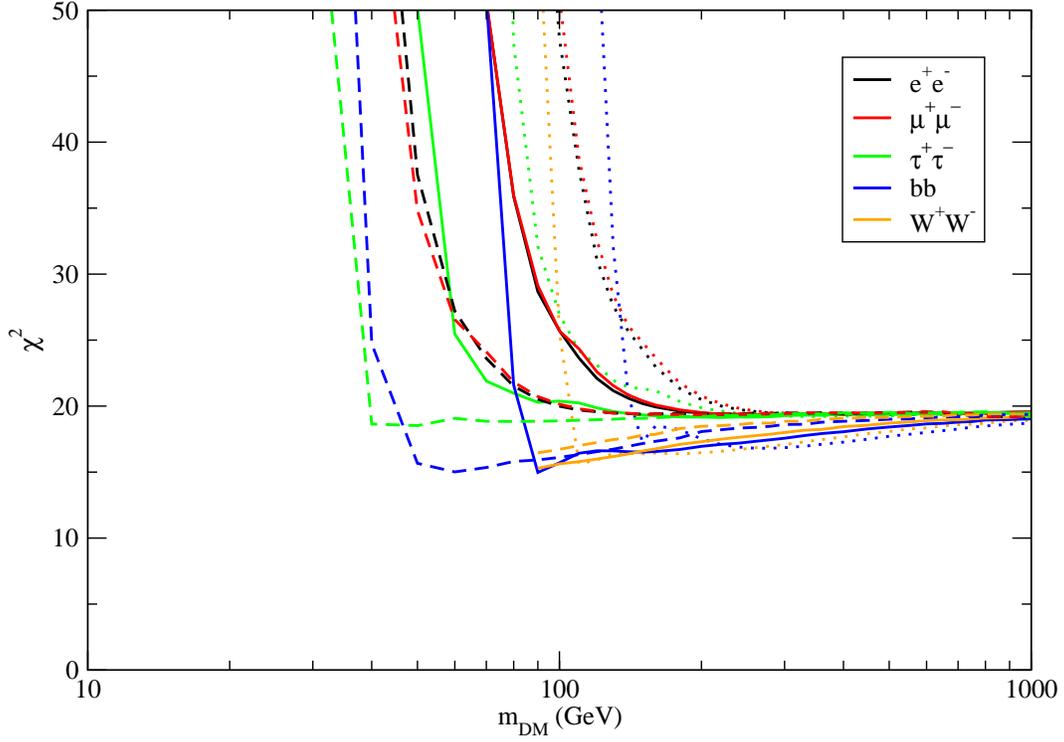}
\caption{The values of $\chi^2$ against $m_{DM}$ for different annihilation channels. Here, the dotted lines, solid lines and dashed lines represent the $\chi^2$ with the $1\sigma$ upper bounds, mean values and $1\sigma$ lower bounds of the dark matter density parameters ($\rho_0$, $r_0$ and $r_{0,\rm NFW}$) respectively. We have assumed $D_0=10^{28}$ cm$^2$ s$^{-1}$.}
\label{Fig4}
\end{center}
\end{figure}

\begin{figure}
\begin{center}
\includegraphics[width=140mm]{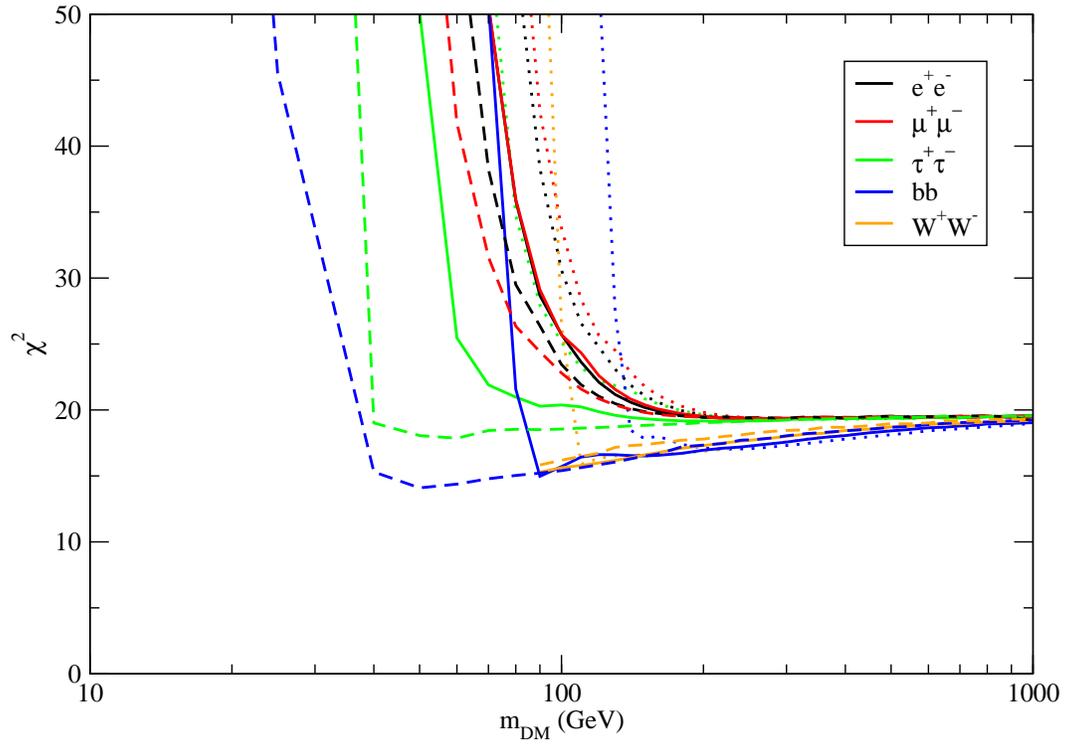}
\caption{The values of $\chi^2$ against $m_{DM}$ for different annihilation channels. Here, the dotted lines, solid lines and dashed lines represent the $\chi^2$ with $B=8$ $\mu$G, $B=4.3$ $\mu$G and $B=2$ $\mu$G respectively. We have assumed $D_0=10^{28}$ cm$^2$ s$^{-1}$.}
\label{Fig5}
\end{center}
\end{figure}

\section{Acknowledgements}
We thank the anonymous referee for useful constructive feedbacks and comments. The work described in this paper was partially supported by the Seed Funding Grant (RG 68/2020-2021R) and the Dean's Research Fund of the Faculty of Liberal Arts and Social Sciences, The Education University of Hong Kong, Hong Kong Special Administrative Region, China (Project No.: FLASS/DRF 04628).

\end{document}